\documentclass[twocolumn,aps,prb,showpacs,amsmath,floatfix,
amssymb,superscriptaddress]{revtex4}

\usepackage{graphicx}
\usepackage{dcolumn}
\usepackage{bm}

\begin{document}

\title{Competition and coexistence of antiferromagnetism and
superconductivity in $R$Ba$_2$Cu$_3$O$_{6+x}$ ($R$$\,=\,$Lu, Y) single
crystals}

\author{A. N. Lavrov}
\affiliation{Institute of Inorganic Chemistry, Lavrentyeva-3,
Novosibirsk 630090, Russia}
\author{L. P. Kozeeva}
\affiliation{Institute of Inorganic Chemistry, Lavrentyeva-3,
Novosibirsk 630090, Russia}
\author{M. R. Trunin}
\affiliation{Institute of Solid State Physics, Institutskaya-2,
Chernogolovka 142432, Moscow region, Russia}
\author{V. N. Zverev}
\affiliation{Institute of Solid State Physics, Institutskaya-2,
Chernogolovka 142432, Moscow region, Russia}

\date{\today}

\begin{abstract}

We use $c$-axis resistivity and magnetoresistance measurements to study the
interplay between antiferromagnetic (AF) and superconducting (SC) ordering
in underdoped $R$Ba$_2$Cu$_3$O$_{6+x}$ ($R$$\,=\,$Lu, Y) single crystals.
Both orders are found to emerge from an anisotropic 3D metallic state, upon
which antiferromagnetism opposes superconductivity by driving the doped
holes towards localization. Despite the competition, the superconductivity
sets in before the AF order is completely destroyed and coexists with
latter in a certain range of hole doping. We find also that strong magnetic
fields affect the AF-SC interplay by both suppressing the superconductivity
and stabilizing the N\'{e}el order.

\end{abstract}


\maketitle

\section{Introduction}

In high-$T_c$ cuprates, the conducting state develops gradually as the
parent antiferromagnetic (AF) insulator is doped with charge
carriers.\cite{RMP_LSCO, mobility, Fermi_arc} At a certain point of this
transformation the superconductivity (SC) sets in, which happens long
before such remnants of the AF insulating state as AF correlations and a
pseudogap in the electron density of states are completely wiped
out.\cite{RMP_LSCO, mobility, Fermi_arc, YBCO_AF} Despite the decades of
intensive research, however, detail mechanisms of the insulator-metal
transformation and, in particular, the role of AF correlations therein
remain unclear. From the experimental point of view, the difficulty stems
from the very strong exchange interaction in CuO$_2$ planes, $J\,$$\sim
\,$$0.1\,$eV,\cite{RMP_LSCO, YBCO_AF} which makes the AF correlations
robust to high temperatures and magnetic fields and leaves no means of
``switching'' them off. For compositions on either side of the
insulator-superconductor transformation, two-dimensional (2D) AF
correlations nucleate in CuO$_2$ planes already at high temperatures
$T\,$$\sim \,$$J/k_B$. As the temperature decreases, the spin correlations
smoothly develop until---depending on the doping level---establishing a
long-range AF order at the N\'{e}el temperature $T_N$, forming a static of
fluctuating ``stripe'' structure, or freezing into a ``cluster spin
glass''.\cite{YBCO_AF, RMP_LSCO, SG, NdSr, Kivelson} This evolution being
spread over a wide temperature range makes it hard to clarify whether and
how the spin textures emerging in CuO$_2$ planes affect the charge motion.

In contrast to the gradual development of 2D AF correlations within the
CuO$_2$ planes, the spin ordering along the $c$ axis---if it takes place
at all---occurs in a dramatically compressed temperature interval. Owing
to the strong anisotropy of exchange interactions, the $c$-axis spin
correlation length $\xi^{AF}_c$ stays shorter than the interplane
distance until the 3D ordering temperature $T_N$ is closely
approached.\cite{YBCO_AF, RMP_LSCO} Only in the vicinity of $T_N$,
$\xi^{AF}_c$ reaches the unit-cell size, causing a 2D--3D crossover in
AF correlations, and then diverges sharply. In fact, it is this abrupt
spin ordering along the $c$ direction that allows the AF transition to
show up in macroscopic properties of lightly doped cuprates. In
$R$Ba$_2$Cu$_3$O$_{6+x}$ crystals ($R$ is a rare earth element), for
instance, the $c$-axis resistivity $\rho_c(T)$ exhibits a steep increase
below $T_N$, though $\rho_{ab}(T)$ stays perfectly smooth through
$T_N$.\cite{AF_SC, anis, ourmag} Hence, although AF correlations within
CuO$_2$ planes are hard to affect, the abrupt AF ordering along the $c$
axis opens a possibility to compare the electron transport {\it with}
and completely {\it without} spin order along one of the directions,
thus clarifying the transport mechanisms and the interplay between
magnetism and superconductivity.

In this paper, we report a $c$-axis magnetoresistance (MR) study of
LuBa$_2$Cu$_3$O$_{6+x}$ (Lu-123) and YBa$_2$Cu$_3$O$_{6+x}$ (Y-123)
single crystals with doping levels spanning the range of the AF-SC
transformation. Sharp AF transitions observed in crystals even with low
$T_N\,$$\approx$$\,20\,$K have allowed us to map precisely the phase
diagram and to single out the impact of spin ordering on the $c$-axis
charge transport. We find that in the doping region of the AF-SC
transformation the (Lu,Y)-123 crystals exhibit a 3D metallic conduction
until the SC or AF ordering intervenes at low temperatures. The N\'{e}el
ordering appears to compete with SC by strongly inhibiting the charge
motion, yet nevertheless the two orders coexist in a certain range of
hole doping. Interestingly, the magnetic field is found to interfere
with both orders, suppressing the superconductivity and stabilizing the
static AF order.

\section{Experimental details}

$R$Ba$_2$Cu$_3$O$_{6+x}$ single crystals with non-magnetic rare-earth
elements $R$$\,=\,$Lu and Y were grown by the flux method and their
oxygen stoichiometry was tuned to the required level by subsequent
high-temperature annealing and quenching.\cite{AF_SC, whisk} Given the
complex distribution of structural defects along the $c$ axis in
flux-grown crystals,\cite{whisk} part of the crystals were cut and
polished to select better pieces for measurements. We confirmed also
that no oxygen-enriched layer was formed at the crystal surface upon
quenching; namely, one of the crystals was dissolved in acid in several
steps, each step being preceded with resistivity measurements, yet no
considerable change in the SC transition was observed.

After the doping level of $R$Ba$_2$Cu$_3$O$_{6+x}$ crystals was roughly
set by changing the oxygen content $x$, the hole density in CuO$_2$
planes $n_h$ was further reversibly tuned using the well studied
phenomenon of chain-layer oxygen ordering: heating the crystal above
100-120$^o$C with subsequent quenching reduced the hole density in
CuO$_2$ planes by up to $20\,\%$, while room-temperature aging gradually
restored it back without any change in stoichiometry.\cite{AF_SC, anis}
For characterizing the doping state of crystals we have selected the
in-plane conductivity $\sigma_{ab}(280\,$K$)$. The latter is a good
measure of the hole density given that the hole mobility stays almost
constant in the doping region under investigation.\cite{mobility} The
particular temperature $T\,$$=\,$$280\,$K is taken to be high enough to
reduce the impact of impurity scattering, yet sufficiently low to avoid
oxygen ordering, and thus change in doping, on the time scale of
measurements. It is worth noting that although the in-plane and the
$c$-axis transport measurements were carried out on separate samples,
the samples were cut from the same crystal, and then annealed and
handled together to provide an exact match in their states.

The $c$-axis resistivity $\rho_c(T)$ was measured using the ac
four-probe technique with a special care taken to avoid mixing of
$\rho_c$ with the in-plane resistivity $\rho_{ab}$. To provide a
homogeneous current flow along the $c$-axis, two current contacts were
painted to almost completely cover the opposing $ab$-faces of the
crystal, while two voltage contacts were placed in small windows
reserved in the current ones.\cite{AF_SC, anis, ourmag}
Magnetoresistance measurements were carried out either by sweeping
temperature at fixed magnetic fields ${\bf H}$$\,\parallel\,$${\bf c}$
up to $16.5\,$T or by sweeping the field at fixed temperature.

\section{Results}
\subsection{$n_h$-$T$-$H$ phase diagram}

\begin{figure}[!t]
\includegraphics*[width=8.2cm]{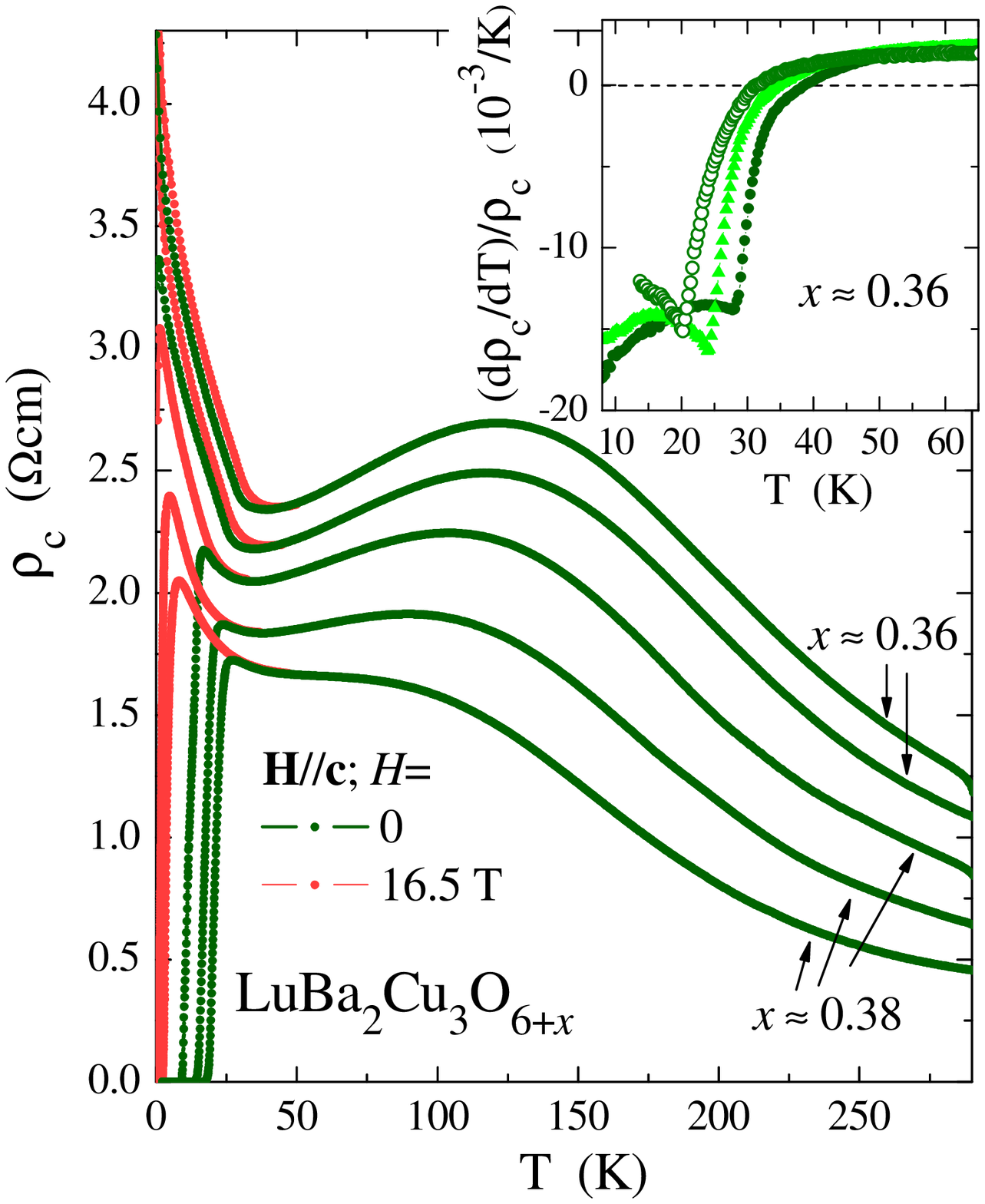}
\caption{(Color online) Out-of-plane resistivity, $\rho_c(T)$, of a
LuBa$_2$Cu$_3$O$_{6+x}$ single crystal at several hole-doping levels
near the AF-SC transformation; the hole doping was tuned by changing the
oxygen content $x$ and the degree of oxygen ordering (several states for
each $x$). The data were taken at zero and 16.5-T magnetic field applied
along the $c$ axis. Inset: An anomaly in the normalized derivative
$(d\rho_c/dT)/\rho_c$ associated with the N\'{e}el transition;
measurements were done after quenching the crystal ($\bullet$) as well
as after room-temperature aging for several hours ($\blacktriangle$) and
days ($\circ$).}
\label{fig1}
\end{figure}

In $R$Ba$_2$Cu$_3$O$_{6+x}$, the hole density in CuO$_2$ planes, $n_h$, can
be tuned continuously by changing the degree of oxygen ordering (see
Sec.~II). This provides a number of states with different $n_h$ for a given
oxygen content $x$, thereby allowing critical regions of the phase diagram
to be studied in detail. We have used this approach to trace how the
$c$-axis resistivity of LuBa$_2$Cu$_3$O$_{6+x}$ single crystals evolves
with the hole density in the narrow doping region where superconductivity
sets in (Fig.~1). As can be seen, the $\rho_c(T)$ curves in this heavily
underdoped region demonstrate a complicated behavior with a
crossover\cite{cross} at 100-150 K. At temperatures above the crossover,
the $c$-axis resistivity smoothly decreases with increasing hole density,
and no peculiarity shows up in this smooth evolution at the doping level of
the SC onset. The low-$T$ resistivity behavior, however, changes
remarkably. From the set of {\it zero-field} $\rho_c(T)$ curves in Fig.~1
one can immediately infer that, upon cooling, the crystal either turns into
the SC state, or instead exhibits a steep increase in the resistivity. This
steep growth in $\rho_c(T)$ with a characteristic asymmetric negative peak
in the derivative (inset in Fig.~1) is a hallmark of the N\'{e}el
transition in $R$Ba$_2$Cu$_3$O$_{6+x}$.\cite{AF_SC, anis, ourmag} The
evolution of $\rho_c(T)$ curves looks therefore like AF and SC are
competing for carriers and the system falls into one of two possible
states, either the insulating AF state or the SC one. It is worth
emphasizing that sharp AF transitions are observed down to remarkably low
$T_N\,$$\approx$$\,20\,$K (inset in Fig.~1) instead of the freezing into a
spin glass usually expected at such temperatures \cite{SG}; the
superconductivity in Lu-123 thus develops directly from the AF state
without any intervening spin-glass region.

\begin{figure*}[!t]
\includegraphics*[width=11cm]{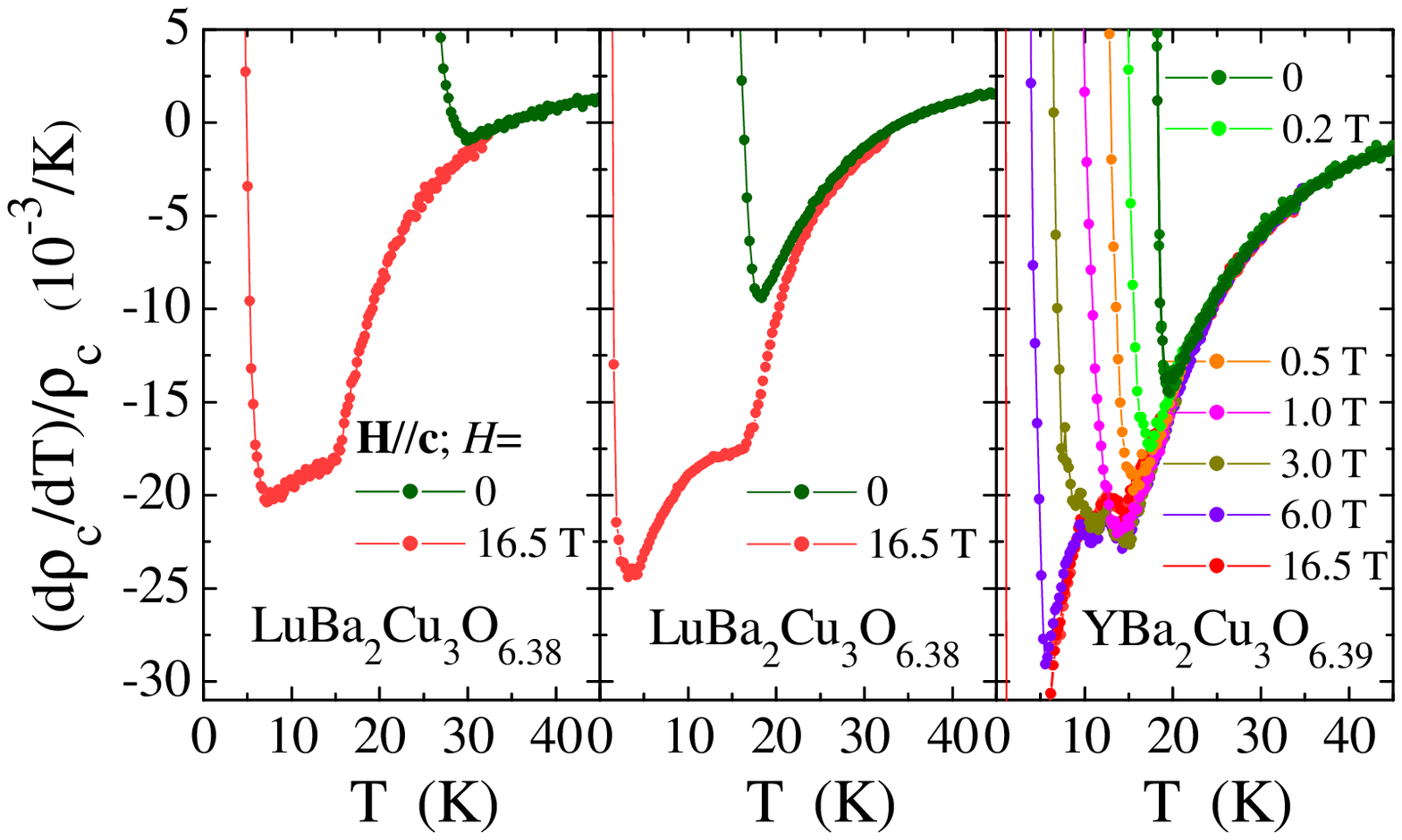}
\caption{(Color online) Suppression of the superconductivity with the magnetic field
in Lu-123 and Y-123 crystals. The unveiled normal-state resistivity
exhibits an anomaly associated with the N\'{e}el transition (compare with
the inset in Fig.1).}
\end{figure*}

One may wonder whether a strong magnetic field will suppress both the AF
and SC orders and create some intermediate metallic state. We find,
surprisingly, that upon suppressing the superconductivity the field
${\bf H}$$\,\parallel\,$${\bf c}$ also recovers the steep increase in
$\rho_c$ associated with the AF ordering (Fig.~1). A study of other
crystals has confirmed that as long as the SC is weak enough to be wiped
out with the 16.5-T field, be that in Lu-123 or Y-123 crystals, the
obtained normal-state $\rho_c(T)$ curves demonstrate clear anomalies at
15-20$\,$K associated with the N\'{e}el transition (Fig.~2). This
behavior indicates that either the AF and SC orders coexist and the
magnetic field simply unveils the resistivity anomaly at $T_N$ hidden by
superconductivity, or the AF order is frustrated in the SC state but
revives as the superconductivity is destroyed.

Although the N\'{e}el transitions in Figs.~1 and 2 look rather sharp for
transition temperatures reduced by 20-25 times from the original
$T_{N0}\,$$\approx \,$410-420~K in parent
$R$Ba$_2$Cu$_3$O$_{6}$,\cite{YBCO_AF} their width together with
insufficient understanding of the interlayer charge transport still
introduce a considerable uncertainty in the $T_N$ evaluation. Indeed,
depending on the mechanisms underlying the observed $c$-axis resistivity
anomalies, $T_N$ should be taken at the position of the negative peak in
the derivative $d\rho_c/dT$ or at the middle of the ``drop'' in
$d\rho_c/dT$ preceding the peak. These two definitions provide $T_N$
values that differ by several K for the curves shown in Figs.~1 and 2.
Nevertheless, whichever approach in evaluating $T_N$ is selected, the
qualitative conclusion on the long-range AF order with $T_N\,$$\approx
\,$15-20$\,$K showing up in low-$T_c$ samples (Fig.~2) remains
unchanged.

\begin{figure}[!b]
\includegraphics*[width=8.65cm]{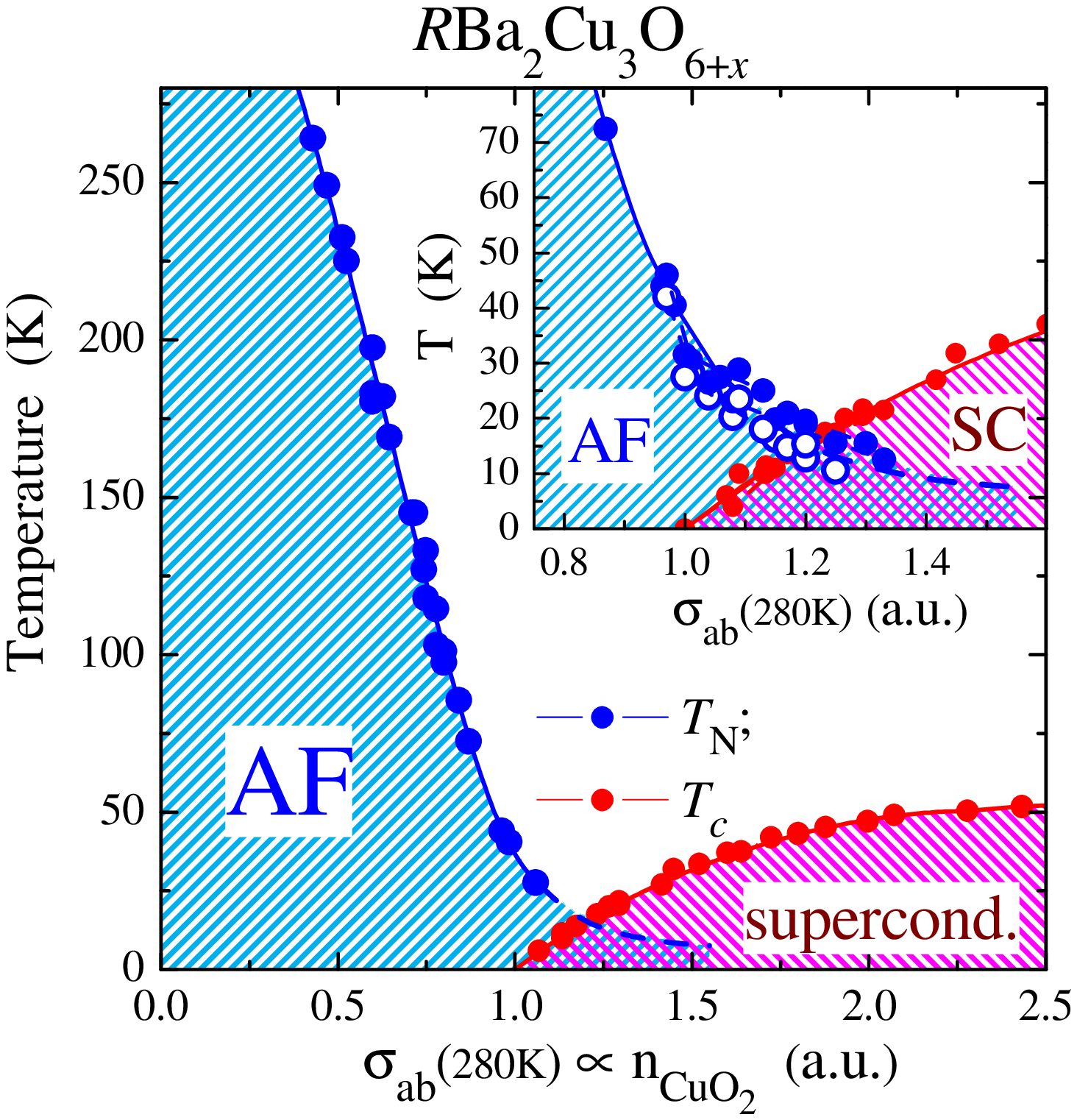}
\caption{(Color online) A phase diagram of $R$Ba$_2$Cu$_3$O$_{6+x}$
($R$ = Lu, Tm, Y) crystals presenting the AF and SC transition
temperatures, $T_N$ and $T_c$, as a function of the in-plane
conductivity $\sigma_{ab}(280 K)$ that is roughly proportional to the
hole density. (To avoid uncertainty associated with the samples'
geometry, $\sigma_{ab}(280 K)$ was normalized by its value at the SC
onset). The critical region of the phase diagram is shown in the inset.
The N\'{e}el temperature was determined either by the position of the
negative peak in the derivative $d\rho_c/dT$ (open circles), or at the
middle of the drop in the derivative (solid circles). Note that in SC
samples the $T_N$ values were determined after suppressing the
superconductivity with 16.5-T field ${\bf H}$$\,\parallel\,$${\bf c}$.}
\end{figure}

The phase diagram in Fig.~3 presenting $T_c$ and $T_N$ as a function of the
in-plane conductivity $\sigma_{ab}(280 K)$ (see Sec.~II) makes the overlap
of the AF and SC regions in $R$Ba$_2$Cu$_3$O$_{6+x}$ apparent. The main
panel in Fig.~3 depicts a general view of the AF-SC transformation
established for several crystals systematically driven through the entire
doping range (most data are taken from our previous studies\cite{AF_SC,
anis}), while the enlarged view of the critical region with new data points
is shown in the inset. As can be seen, the AF phase boundary does not
terminate at the onset of superconductivity. Instead, it extends at least
up to 30\% higher doping, and crosses the $T_c$ line at $\approx
\,$$15\,$K, in contrast to another prototype high-$T_c$ cuprate
La$_{2-x}$Sr$_x$CuO$_4$ where the long-range AF order is destroyed long
before the SC sets in.\cite{RMP_LSCO, SG}

\begin{figure}[!t]
\includegraphics*[width=8.65cm]{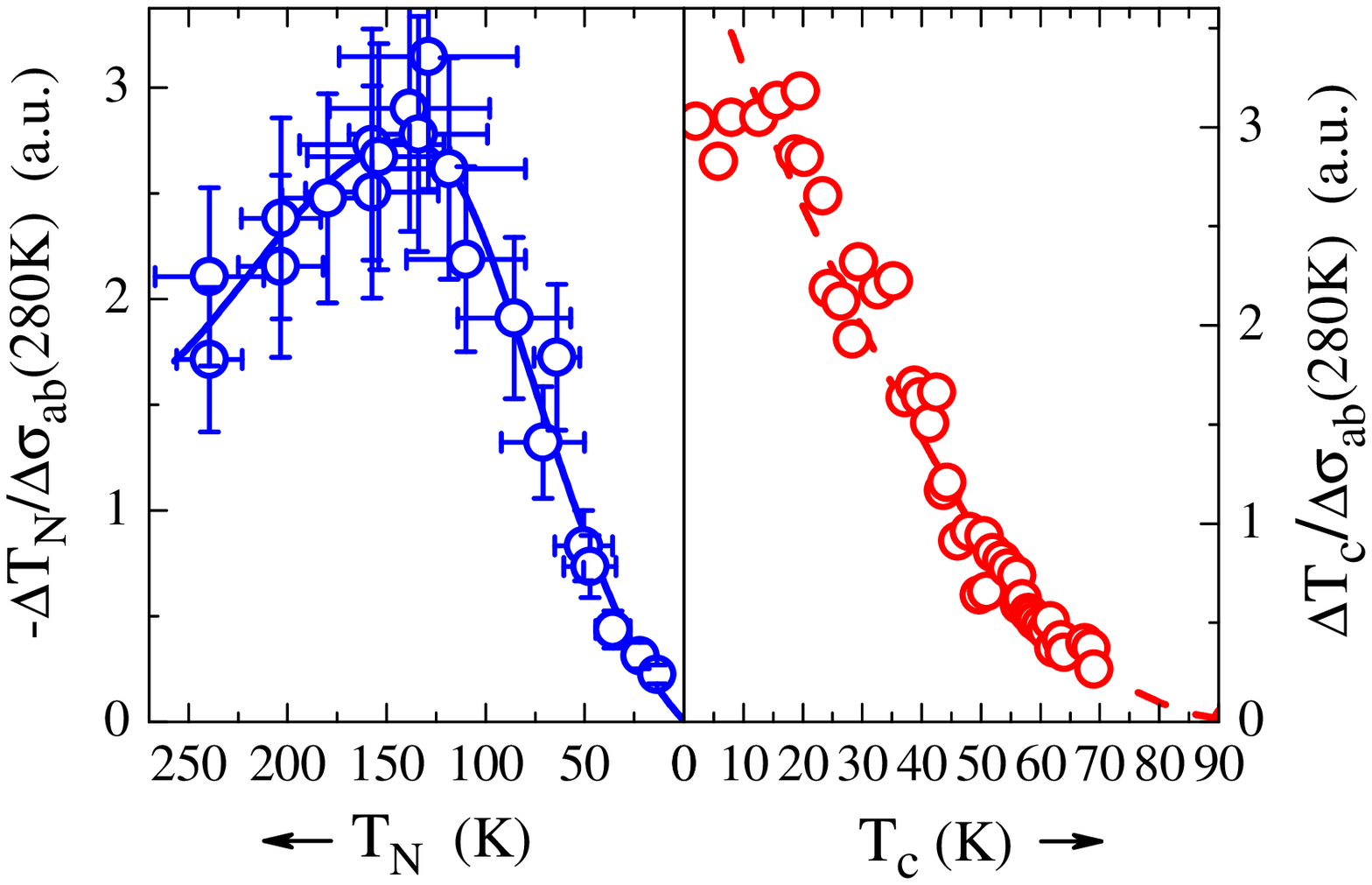}
\caption{(Color online) Shifts of $T_N$ and $T_c$ normalized by
corresponding changes in $\sigma_{ab}$ as observed upon room-temperature
aging of $R$Ba$_2$Cu$_3$O$_{6+x}$ crystals. The data are presented as a
function of $T_N$ and $T_c$. The horizontal bars in the left panel
indicate the actual interval of the $T_N$ shift in each experiment.}
\label{fig4}
\end{figure}

An important issue is the reliability of the presented phase diagram,
especially given the controversy remaining in publications.
\cite{YBCO_AF, SG, diagram2, diagram3} Indeed, the observed SC and AF
transitions are not very sharp, leaving some room for doubts. Besides,
however sharp the transitions might be, a possibility of a peculiar
macroscopic inhomogeneity imitating the phase coexistence will still
remain. For the oxygen compositions under discussion, we have ruled out
one possible source of inhomogeneity associated with a phase separation
into oxygen-rich and oxygen-poor domains: The behavior of Lu-123
crystals quenched from high temperatures, and thus possessing
homogeneous oxygen distribution, showed no qualitative difference from
that of crystals aged at room temperature for an ultimate 10-years
period.

A static cation inhomogeneity over the crystal as a source of the AF and SC
phase coexistence is, however, hard to rule out by conventional means.
Fortunately, in R-123 crystals one has an additional tool to check the
phase diagram by employing the hole doping though the oxygen
ordering.\cite{AF_SC, anis} In fact, by aging the crystals at room
temperature and measuring their conductivity $\sigma_{ab}(280 K)$, $T_c$,
and $T_N$ as a function of time, one can evaluate the slops of phase
boundaries in the diagram shown in Fig.~3. An important point here is that
the relative changes in $T_c$ and $T_N$ observed upon aging are insensitive
to static macroscopic inhomogeneities. The phase-boundary slopes measured
in the described way are presented in Fig.~4 as a function of $T_c$ and
$T_N$; this way of the data presentation is selected to avoid any
uncertainty related to ill-defined parameters such as the hole density. As
expected, the slope of the SC phase boundary, $\Delta T_c/\Delta
\sigma_{ab}(280 K)$, gradually increases with decreasing $T_c$ and the
superconductivity disappears abruptly at a critical doping. The behavior of
the AF boundary is much different; initially it indeed becomes steeper with
increasing doping and decreasing $T_N$, but then the slop of the $T_N$ line
decreases dramatically, tending to zero at zero $T_N$. This flattening
unambiguously confirms the intrinsic origin of the AF-order tail entering
the SC region in the phase diagram in Fig.~3.

\begin{figure}[!t]
\includegraphics*[width=7.7cm]{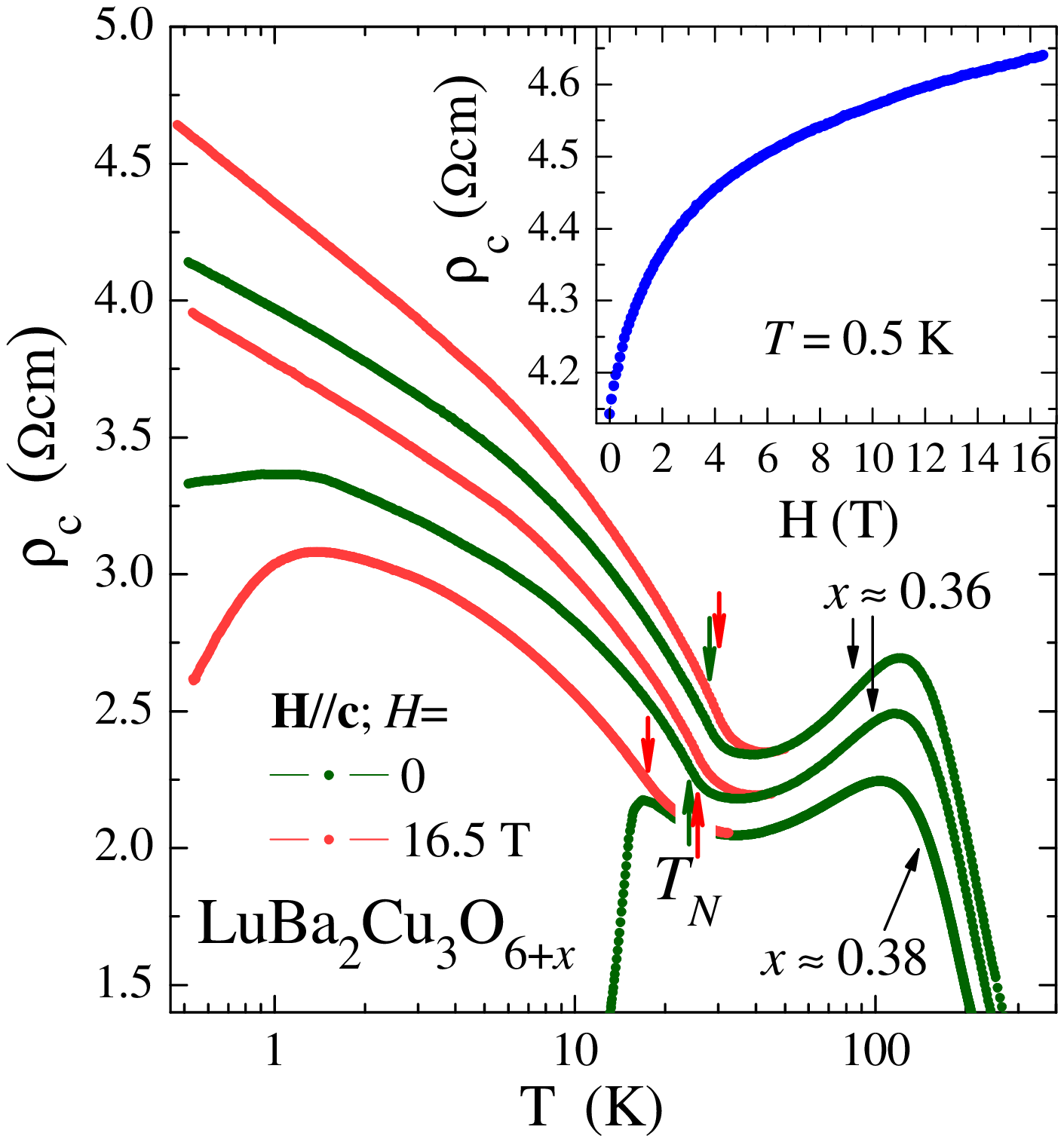}
\caption{(Color online) Semilogarithmic plot of $\rho_c(T)$ illustrating the
additional low-$T$ resistivity caused by the AF ordering and the impact
of the magnetic field ${\bf H}$$\,\parallel\,$${\bf c}$ on both $T_N$
and $\rho_c$. Two sets of curves for $x$$\,\approx \,$0.36 correspond
to the quenched and partially aged states of the crystal. Low-$T$
magnetoresistance data for the quenched $x$$\,\approx \,$0.36 sample
are shown in the inset.}
\label{fig5}
\end{figure}

What still remains to be clarified is whether the AF and SC orders
actually coexist in their overlapping region on the phase diagram, or
the AF order is restored by the magnetic field upon destroying the
superconductivity. The latter is not completely impossible given that
the magnetic field is found to inhibit the hole motion and to stabilize
the N\'{e}el order even in non-SC $R$Ba$_2$Cu$_3$O$_{6+x}$ crystals
(Fig.~5). One can see that the N\'{e}el temperature indicated in the
plot by an arrow, increases by $\sim\,$2-3~K under 16.5-T field ${\bf
H}$$\,\parallel\,$${\bf c}$. As a result, the corresponding resistivity
upturn also starts from higher temperatures, which causes a significant
magnetoresistance $\Delta \rho_c/\rho_c$ reaching $\approx$$\,12\%$ at
low $T$ (inset in Fig.~5). It is worth recalling that the resistivity
measurements are incapable of providing information on $T_N$ values
within the SC state and the data points in Fig.~3 located inside the SC
region were obtained after the SC was killed by the 16.5-T magnetic
field. If the field-induced increase of $T_N$ in the SC region is
two-three times larger than in non-SC ones, the true magnetic state
underlying the superconductivity in $R$-123 may well be a ``cluster spin
glass'' which is driven to a long-range AF order only with the SC
suppression.

\subsection{AF order and the $c$-axis conductivity}

\begin{figure}[!t]
\includegraphics*[width=8.65cm]{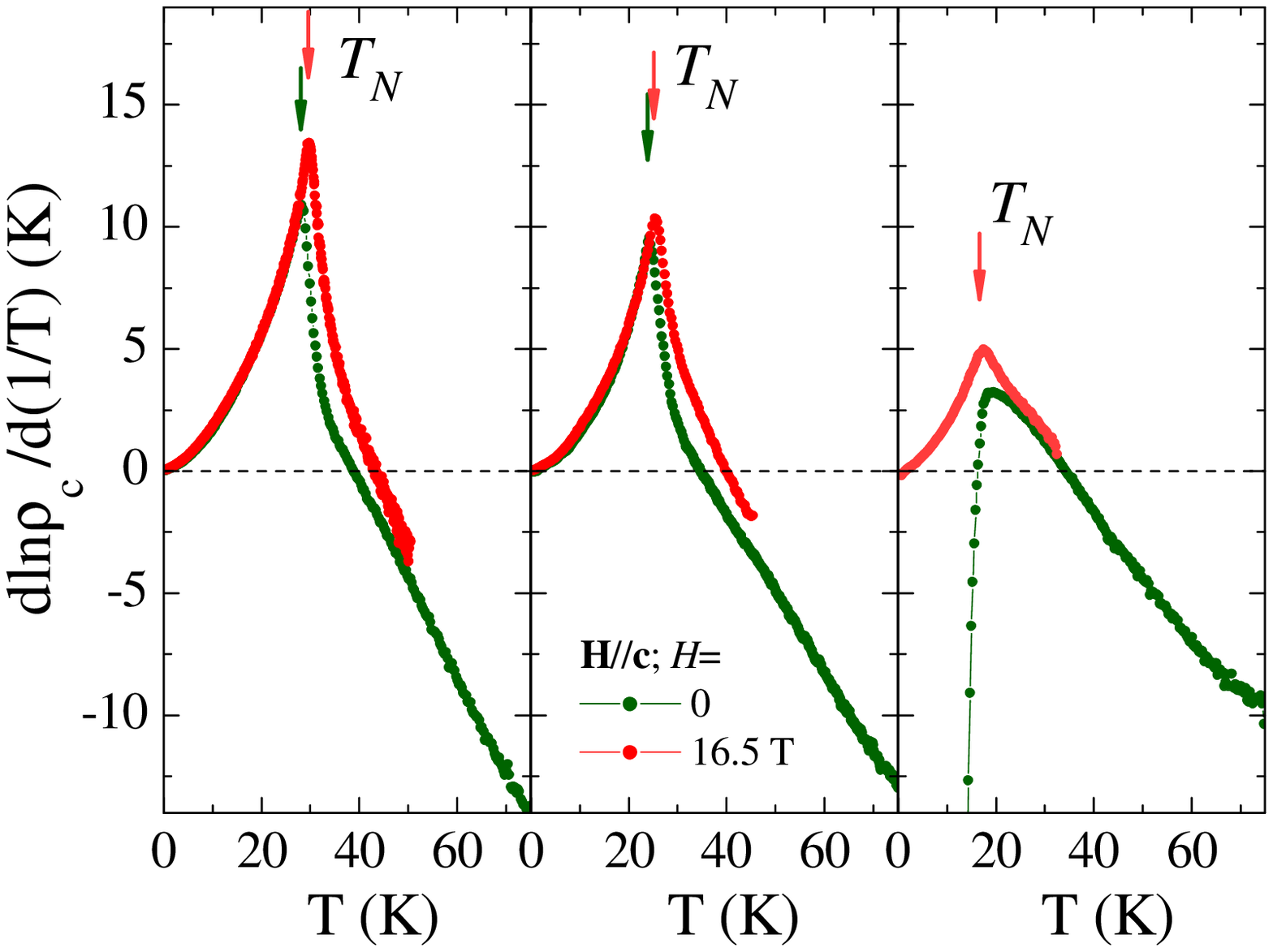}
\caption{(Color online) Evolution of the derivative $d\ln\rho_c/d(1/T)$ upon
changing the N\'{e}el temperature. The data were obtained on the same
LuBa$_2$Cu$_3$O$_{6+x}$ crystal at $x$$\,\approx \,$0.36, in the quenched
(left panel) and partially aged (middle panel) states, and at $x$$\,\approx
\,$0.38 in the quenched state (right panel).}
\label{fig6}
\end{figure}

After establishing the phase diagram, we can consider what is the
``normal state'' (at $T\,$$>\,$$T_N$, $T_c$) that both the AF and SC
orders originate from, and how it is affected by the AF ordering. At a
first glance, the long-range antiferromagnetism just localizes the doped
holes, thus strictly opposing the superconductivity (Figs.~1 and 5).
However, there exists a narrow doping range on the phase diagram in
Fig.~3 where the superconductivity sets in at $T\,$$<\,$$T_N$ (see also
Figs.~1 and 2), which raises additional questions concerning the
mechanism of hole transport in the AF state. In particular, does the
$c$-axis transport actually acquire an activated character with
establishing the long-range AF order? If the AF-ordered state is
inevitably an insulator with an appreciable energy gap, how does the
superconductivity emerge at $T_c\,$$<\,$$T_N$?

The simplest way of evaluating the activation energy is to plot the
derivative $d\ln\rho_c/d(1/T)=\Delta(T)$ which gives a $T$-independent
$\Delta$ for ordinary activation behavior $\rho=A\exp(\Delta /T)$ or an
effective gap $\Delta(T)\propto T^{1-k}$ for variable range hopping
$\rho=A\exp(B/T^k)$. In contrast to expectations, the derivative data in
Fig.~6 show a clear positive curvature ($k\,$$<\,$0) in the entire
temperature region below $T_N$, which is inconsistent with a simple
activated transport. The AF ordering therefore inhibits the hole motion,
causing a sharp kink at $T_N$, but does not immediately turn the system
into an insulator. With further reduction of the carrier density and
increase of $T_N$ above 50-100~K, the hole motion indeed switches to the
hopping regime,\cite{ourmag} yet in the low-$T_N$ region shown in
Figs.~5 and 6 the quasi-metallic conduction survives, leaving some room
for superconductivity. Strictly speaking, a possibility that the
resistivity even in low-$T_N$ crystals in Figs.~5 and 6 will acquire an
insulating behavior in the sub-Kelvin region still remains, but the
corresponding insulating gap in that case would be too small to affect
the occurrence of superconductivity.

The derivative plots in Fig.~6 have other important implications. First,
all the derivatives $d\ln\rho_c/d(1/T)$ at low $T$ appear to follow
roughly the same curve until breaking down at $T_N$. The same low-$T$
resistivity behavior is observed even in superconducting crystals after
suppressing SC with magnetic field (right panel). The $c$-axis transport
mechanism in the spin ordered state is therefore kept unchanged in the
critical doping range of the AF-SC transformation; what changes with
doping and magnetic field is merely the N\'{e}el temperature (Fig.~6).
Second, the $\rho_c(T)$ upturn at low $T$---marked by the positive sign
of the derivative in Fig.~6---is solely, or at least predominantly
caused by the AF ordering. As the spin order melts above $T_N$, the
resistivity quickly switches to the metallic behavior [negative sign of
$d\ln\rho_c/d(1/T)$]. With decreasing $T_N$ the region of the
resistivity upturn shrinks with a clear tendency to vanish completely
for $T_N\rightarrow 0$ (Fig.~6). Therefore, in the $T_N=0$ limit (if it
could be reached without getting into the SC state) the $c$-axis
resistivity would keep the metal-like behavior down to low $T$, slowly
decreasing and approaching a constant value at zero temperature.

The obtained data indicate that, in the absence of the interplane AF order,
the charge carriers in $R$Ba$_2$Cu$_3$O$_{6+x}$ can move along the $c$ axis
without thermal activation. Regardless of whether the non-activated
interlayer conduction occurs through a coherent, metallic transport or by
an energy-conserving tunneling, it requires itinerant quasiparticles to be
already formed in CuO$_2$ planes. The latter in turn implies the in-plane
charge transport in the non-AF-ordered state to be invariably metallic as
well. Indeed, the in-plane resistivity $\rho_{ab}(T)$ was reported to
exhibit a metallic behavior without any upturn until entering the N\'{e}el
state.\cite{AF_SC, anis, Trunin} Consequently, the ``normal state'' playing
a role of a starting point for the AF-SC phase competition in
$R$Ba$_2$Cu$_3$O$_{6+x}$ is an anisotropic 3D metal that experiences the
impact of SC and AF ordering upon lowering the temperature.

\section{Discussion}

By now, many theoretical approaches have been suggested to describe the
frustration that the doped holes introduce to the spin system and the
resulting suppression of the long-range AF order.\cite{Kivelson, Aharony,
Nagaev, spin_vor, spiral1, spiral2, Gooding} Namely, the doped holes were
proposed to switch the exchange interaction between the nearest Cu spins
from antiferro- to ferromagnetic while being localized,\cite{Aharony} or to
continuously break the AF bonds while moving by the nearest-neighbor
hopping.\cite{Kivelson, Nagaev} Spin distortions accompanying the doped
holes are not restricted to the nearest Cu ions, but spread for several
unit cells away and may form spin polarons, 2D spin
vortices,\cite{spin_vor} spin spirals,\cite{spiral1, spiral2} AF domain
walls,\cite{Kivelson} or complex spin textures controlled by the
distribution of the dopants.\cite{Gooding} Whatever the case, upon
increasing the hole density, the resulting spin distortions are suggested
to gradually reduce the average staggered magnetization in each CuO$_2$
plane to zero, thereby making impossible the interlayer spin ordering and
hence the long-range AF order.

While the mentioned above theoretical approaches have been positively
tested on the prototype high-$T_c$ cuprate La$_{2-x}$Sr$_x$CuO$_4$, the
considerably different behavior of $R$Ba$_2$Cu$_3$O$_{6+x}$ may call for
reconsidering the impact of doped holes on the long-range AF order. Indeed,
the AF state in $R$Ba$_2$Cu$_3$O$_{6+x}$ survives up to 2-3 times higher
hole concentrations than in La$_{2-x}$Sr$_x$CuO$_4$, pointing to a
significant material dependence of the AF-order suppression. Furthermore,
the low-$T_N$ tail of the AF order spreading in the phase diagram towards
SC compositions (Fig.~3) is hard to explain in the framework of those
approaches, since they predict no slowing down of the spin frustration with
increasing doping. One more difficulty for the existing models is the sharp
AF transitions observed down to low $T_N$'s, which are hard to reconcile
with strongly frustrated spin states and disordered spin textures in
CuO$_2$ planes suggested to emerge with doping.\cite{Aharony, spin_vor,
Gooding}

Nevertheless, most of the observed features may be qualitatively accounted
for by assuming the doped holes to break the long-range AF order in CuO$_2$
planes into 2D domains. \cite{Kivelson} Indeed, since the N\'{e}el
temperature in that case should be determined by the interlayer spin
coupling integrated over an average domain,\cite{YBCO_AF, RMP_LSCO} a
gradual reduction of the domain size with doping can explain why $T_N$
approaches zero value slowly and smoothly instead of vanishing abruptly
(Fig.~3). It is worth recalling that the interlayer coupling in
$R$Ba$_2$Cu$_3$O$_{6+x}$ is not frustrated \cite{YBCO_AF} in contrast to
La$_{2-x}$Sr$_x$CuO$_4$,\cite{RMP_LSCO} and thus should allow the N\'{e}el
order in the former to survive up to much higher doping.

The idea of moving holes frustrating the AF order implies a reverse impact
as well, which makes the origin of the observed antiferromagnetism-induced
hole localization intuitively clear. As the penalty for frustrating the AF
bonds exceeds the kinetic energy of a doped hole, the hole should get
trapped by the environment. While in CuO$_2$ planes the hole hopping to the
nearest-neighbor sites is inhibited smoothly \cite{mobility} due to the
gradual development of AF correlations, the suppression of the $c$-axis
hole hopping should occur abruptly with the $c$-axis spin ordering at
$T_N$, exactly as observed experimentally. However a question remains: when
exactly does the charge transport switch its character from metallic to
non-metallic, i.e. thermally-activated hopping? A comparison of the
$c$-axis conductivity for the cases of spins being or being not AF ordered
along that axis has demonstrated that in the absence of spin order the
holes move without thermal activation. Therefore, the ``normal state'' in
the AF-SC transformation region (at temperatures above $T_N$, $T_c$) is
certainly a 3D metal. As the AF order develops, the holes are driven to
localization, yet the exact position of the metal-insulator transition
(MIT) remains unsettled.

Previous attempts to determine the MIT location on the phase diagram of
$R$Ba$_2$Cu$_3$O$_{6+x}$ employed both the charge-transport and
heat-transport measurements, yet resulted in somewhat controversial
conclusions; namely, the MIT was suggested to occur at hole densities
lower, equal, or even higher than the SC onset (see
Refs.~\onlinecite{Our_MIT, Sun_MIT} and references therein). Is there a
universal doping level for MIT in $R$Ba$_2$Cu$_3$O$_{6+x}$ at all? An
important result following from the present study is that the
metal-insulator transformation is largely driven by the AF spin ordering.
Given that the AF correlations in $R$Ba$_2$Cu$_3$O$_{6+x}$ are not
determined solely by the hole density, but develop with decreasing
temperature, get stabilized by magnetic fields and, presumably, depend on
some material parameters, there is little reason to expect the MIT to be
located on the phase diagram at some universal hole density. Instead, the
metal-insulator transformation may take place depending on doping, magnetic
field, material parameters, and, in a sense, even on temperature. The
latter means that the hole localization gains strength on cooling due to
the development of the AF order and thus the metal-insulator border may be
crossed at some transition temperature.\cite{MIT_T}

What still remains to be understood is how the magnetic field enters the
AF-SC interplay. One possibility is that the magnetic field affects both
orders almost independently, suppressing superconductivity and stabilizing
the long-range AF state. Indeed, a magnetic field applied to a 2D
Heisenberg antiferromagnet can introduce an anisotropy to the spin system,
thus suppressing the spin fluctuations and allowing the long-range N\'{e}el
order to be established. Alternatively, the AF and SC ground states may be
separated by a first-order transition, as were predicted, for instance, in
the SO(5) theory, \cite{SO5} and the magnetic field may directly affect
their competition. Microscopic probes capable of detecting the AF order
inside the superconducting state are indispensable for this issue to be
clarified.

\begin{acknowledgments}
We thank V. F. Gantmakher and D. V. Shovkun for fruitful discussions and
acknowledge support by RFBR (grants 05-02-16973 and 09-02-01224) and the
integration project SB RAS No.81.
\end{acknowledgments}


\begin{thebibliography}{99}
\bibitem{cross} A pronounced crossover was reproducibly observed in
the $c$-axis resistivity of underdoped $R$Ba$_2$Cu$_3$O$_{6+x}$
crystals\cite{AF_SC, anis, ourmag, whisk} and some other layered
oxides\cite{Valla}, yet its origin is still insufficiently understood.
\bibitem{MIT_T} In a strict sense MIT is determined only at $T=0$. However,
if the state of the electronic system changes with temperature, one can
introduce an effective MIT temperature, $T_{\text{MIT}}$, so that the
states ``quenched'' from $T>T_{\text{MIT}}$ and $T<T_{\text{MIT}}$ would
correspond to a metal and an insulator at $T=0$.
\end{thebibliography}
\end{document}